\documentclass[aps,pra,twocolumn,amsmath,amssymb,floatfix,nofootinbib]{revtex4}
\usepackage{graphics}      % standard graphics specifications
\usepackage{graphicx}      % alternative graphics specifications
\usepackage{longtable}     % helps with long table options
\usepackage{url}           % for on-line citations
\usepackage{bm}            % special 'bold-math' package
\usepackage[dvips]{epsfig,color}
\usepackage{dcolumn}

\def\bra{\langle}
\def\ket{\rangle}
\def\S{\boldsymbol{S}}
\def \beq{\begin{equation}}
\def \eeq{\end{equation}}
\def \beqarr{\begin{eqnarray}}
\def \eeqarr{\end{eqnarray}}
\def \bspt{\begin{split}}
\def \espt{\end{split}}
\def \bef{\begin{figure}}
\def \enf{\end{figure}}

\newcommand{\abs}[1]{\lvert#1\rvert}

\begin{document}

\title{Block Entanglement Entropy of Ground States with Long-Range Magnetic Order}

\author{Wenxin Ding}
\affiliation{NHMFL and Department of Physics, Florida State
University, Tallahassee, Florida 32306, USA}

\author{Nicholas E. Bonesteel}
\affiliation{NHMFL and Department of Physics, Florida State
University, Tallahassee, Florida 32306, USA}

\author{Kun Yang}
\affiliation{NHMFL and Department of Physics, Florida State
University, Tallahassee, Florida 32306, USA}

\begin{abstract}

In this paper we calculate the block entanglement entropies of
spin models whose ground states have perfect antiferromagnetic or
ferromagnetic long-range order. In the latter case the definition
of entanglement entropy is extended to properly take into account
the ground state degeneracy. We find in both cases the entropy
grows logarithmically with the block size. Implication of our
results on states with general long-range order will be discussed.

\end{abstract}

\date{\today}

\maketitle

%%%%%%%%%%%%%%%%%%%%%%%%%%%%%%%%%%%%%%%
%                                 %%%%%
% Section I: Introduction         %%%%%
%                                 %%%%%
%%%%%%%%%%%%%%%%%%%%%%%%%%%%%%%%%%%%%%%

\section{INTRODUCTION}

Entanglement is the hallmark as well as most counterintuitive
feature of quantum mechanics. Among various ways to quantify
entanglement, bipartite block entanglement entropy emerged as a
concept of central importance in quantum information
science\cite{nielsenchuang}, and has been receiving growing
attention in other branches of physics recently. For example, it
was suggested to be a possible source of black hole
entropy\cite{bombelli,srednicki}. In condensed matter or many-body
physics, the entanglement entropy has been increasingly used as a
very useful and in some cases indispensable way to characterize
phases and phase transitions, especially in strongly correlated
phases\cite{amico}. In this context the most important result is
perhaps the so-called area law\cite{bombelli,srednicki,amico},
which states that in the thermodynamic limit, the entropy should
be proportional to the area of the boundary that divides the
system into two blocks. There are a few very important
examples\cite{amico,wolf,cardy} in which the area law is violated,
most of which involve quantum criticality\cite{cardy}; the
specific manner with which the area law is violated is tied to
certain universal properties of the phase or critical point. In
some other cases, important information about the phase can be
revealed by studying the leading {\em correction} to the area
law\cite{kitaevpreskill,levinwen,fradkinmoore}; for example this
is the case for topologically ordered
phases\cite{kitaevpreskill,levinwen}.

Comparatively there have been relatively few studies of the behavior
of entanglement entropy in states with traditional long-range
order\cite{vidal05,vidal06,vidal07}. This is perhaps because of the expectation that ordered states
can be well described by mean-field theory, and in mean-field theory
the states reduce to simple product states that have no
entanglement. In particular in the limit of perfect long-range order
the mean-field theory becomes ``exact", and the entanglement entropy
should vanish. In this paper we will show that this is {\em not} the
case, and interesting entanglement exists in states with perfect
long-range order. We will study two exactly solvable spin-1/2 models:
(i) An unfrustrated antiferromagnet with infinite range (or constant)
antiferromagnetic (AFM) interaction between spins in opposite
sublattices, and ferromagnetic (FM) interaction between spins in the
same sublattice; (ii) An ordinary spin-1/2 ferromagnet with arbitrary
FM interaction among the spins\cite{note}. While the ground states
have perfect long-range order for both models, we show that they both
have non-zero entanglement entropy that grow logarithmically with the
size of the subsystem.

The paper is organized as follows. In Sec. II we introduce the
model (i) and present its exact solution. In Sec. III we calculate
the reduced density matrix of a subsystem to obtain the exact
expression of the entanglement entropy for model (i). Based on
those results, in Sec. IV we study the scaling behavior of the
entropy as the system size tends to infinity under two different
partition limits and compare it with numerical calculations.
Section V is devoted to model (ii). Finally, in Sec. VI we
summarize and discuss the results of this paper. Some mathematical
definitions, notations and details are given in the appendices.

%%%%%%%%%%% SECTION I END %%%%%%%%%%%%%

%%%%%%%%%%%%%%%%%%%%%%%%%%%%%%%%%%%%%%%
%                                 %%%%%
% Section II: Model, GS & RDM     %%%%%
%                                 %%%%%
%%%%%%%%%%%%%%%%%%%%%%%%%%%%%%%%%%%%%%%

\section{Antiferromagnetic spin MODEL AND GROUND STATE}
\begin{figure}
  % Requires \usepackage{graphicx}
  \includegraphics[width=8cm]{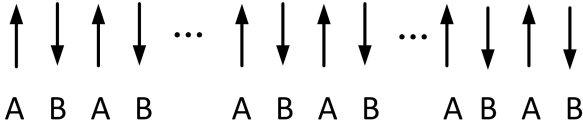}\\
  \caption{Two-sublattice model: two sublattices labeled A and B interpenetrating each other.}\label{Fig.lattice}
\end{figure}

We consider a lattice model composed of two sublattices
interpenetrating each other as in Fig. \ref{Fig.lattice}, with
interaction of infinite range, i.e., every spin interacts with all
the other spins in the system, with interaction strength
independent of the distance between the spins. Within each
sublattice, the interaction is ferromagnetic, and between the
sublattices the interaction is antiferromagnetic; as a result
there is no frustration. The Hamiltonian is written as,
%eq 1
\begin{equation}
\label{Ham}
H=-J_A\sum_{i,j\in A}\S_i \cdot \S_j-J_B\sum_{i,j\in B}\S_i
\cdot \S_j+J_0\sum_{i\in A,j\in B}\S_i \cdot \S_j,
\end{equation}
with $J_A, J_B, J_0 > 0$. The ground state of Eq. (\ref{Ham}) can be
solved in the following manner\cite{yusuf}. Let $$\S_A =
\sum_{i\in A} \S_i,\ \ \ \S_B=\sum_{i\in B}\S_i,\ \ \
\S=\S_A+\S_B,$$ then the Hamiltonian can be written as,
%eq 2
\begin{equation}
\begin{split}
H &= -J_A\S_A^2-J_B\S_B^2+J_0\S_A\cdot\S_B\\
  &= -J_A\S_A^2-J_B\S_B^2 + \frac{J_0}{2}(\S^2-\S_A^2-\S_B^2).\\
\end{split}
\end{equation}
Since $[\S, H]=[\S_A, H]=[\S_B, H]=0$, eigenstates of $H$ can be
chosen to be simultaneous eigenstates of $\S^2$, $\S_A^2$ and
$\S_B^2$, with angular momentum eigenvalues $S$, $S_A$ and $S_B$
respectively, such that the energy eigenvalue is,
%eq3
\begin{equation}
\frac{J_0}{2}S(S+1)-(J_A+\frac{J_0}{2})S_A(S_A+1)-(J_B+\frac{J_0}{2})S_B(S_B+1).
\end{equation}
To minimize the energy, we must first maximize $S_A$ and $S_B$,
and then minimize $S$, i.e., $S = |S_A-S_B|, S_A = \frac{N_A}{2},
S_B = \frac{N_B}{2}$. Here $N_A$ and $N_B$ represent the number of
corresponding sublattice sites. Physically this means that spins
in the same sublattice are all parallel to each other, while spins
in opposite sublattices are antiparallel. Thus the ground state
is,
%eq 4
\begin{equation}
|S_AS_B;Sm\ket = \bigg|\frac{N_A}{2},\frac{N_B}{2};
 \bigg|\frac{N_A}{2} - \frac{N_B}{2}\bigg|, m \bigg\ket.
\end{equation}

%fig.1
%end fig.1

In this paper we only consider the simplest case with $N_A = N_B =
N$, thus the total system size is $2N$. Then the ground state is
reduced to an antiferromagnetic ground state which has zero total
spin,
\beq |S_AS_B;Sm\ket=|\frac{N}{2}\frac{N}{2};00\ket.
\label{singletneel}
\eeq
This ground state has perfect Neel order, as manifested by the
spin-spin correlation function,
\begin{equation}
\begin{split}
&\langle \S_i\cdot\S_j\rangle ={1\over 4},\hskip 0.4 cm i, j \in A \hskip 0.2 cm {\rm or} \hskip 0.2 cm i, j \in B;\\
&\langle \S_i\cdot\S_j\rangle =-{1\over 4}-{1\over 4N}\rightarrow -{1\over 4}, \hskip 0.2 cm i\in A \hskip 0.2 cm {\rm and} \hskip 0.2 cm j \in B.
\end{split}
\end{equation}

%%%%%%%%%%%%%%%%%%%%%%%%%%%%%%%%%%%%%%%%%%%%%%%
%                                         %%%%%
% Section III: Reduced Density Matrix     %%%%%
%                                         %%%%%
%%%%%%%%%%%%%%%%%%%%%%%%%%%%%%%%%%%%%%%%%%%%%%%
\section{THE REDUCED DENSITY MATRIX and entanglement entropy}
\begin{figure}
\includegraphics[width = 8cm]{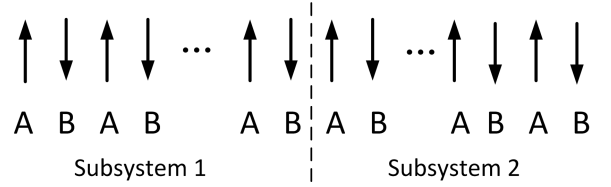}
\caption{Bipartite division of the system. We divide the system
  spatially in to two parts and label them subsystem 1 and subsystem 2,
  respectively. One of the main tasks of this paper is to evaluate the
  entanglement entropy between these two parts.}\label{Fig.division}
\end{figure}

We divide the system spatially into two subsystems which are labeled 1
and 2 respectively, as shown in Fig. \ref{Fig.division}, and study
the ground state entanglement entropy $E$ between these two subsystems.
$E$ is defined to be the von Neumann entropy of either of the
subsystems, $E_{1}$ or $E_{2}$, which can be calculated from the
reduced density matrix,
\begin{equation}
\label{entanglemententropy}
E=E_{1} = E_{2} = -tr(\rho_1 \ln \rho_1) = -tr(\rho_{2} \ln \rho_{2}),
\end{equation}
where $\rho_1$ is the reduced density matrix of the subsystem 1,
obtained from the density matrix $\rho = |\frac{N}{2} \frac{N}{2}
; 00 \ket \bra  \frac{N}{2}\frac{N}{2};00|$ of the whole system by
tracing out degrees of freedom of the other subsystem, $\rho_1 =
tr_{(2)}(\rho)$, and vice versa.

%begin fig2
%end fig2

To solve for the explicit form of the reduced density matrix, we
proceed as follows.
First, we further decompose the system into four parts, $\S_{A_1},\
\S_{A_2},\ \S_{B_1},\ \S_{B_2}$, with
\beq
\begin{cases}
  \S_{A_1} = \sum_{i\in A \wedge i \in 1} \S_i,\ \ \S_{A_2} = \sum_{i\in
  A \wedge i \in 2} \S_i;\\
\S_{B_1} = \sum_{i\in B \wedge i \in 1} \S_i, \ \ \S_{B_2} =
  \sum_{i\in B \wedge i \in 2} \S_i.
\end{cases}
\eeq
Therefore, these operators satisfy the following relations,
\begin{equation}
  \begin{cases}
    \S_A = \S_{A_1}+\S_{A_2},\ \ \ \S_B = \S_{B_1}+\S_{B_2};\\
    \S_1 = \S_{A_1}+\S_{B_1},\ \ \ \S_2 = \S_{A_2}+\S_{B_2}.\\
  \end{cases}
\end{equation}
%%%%%%%%%%%%
%newly added
%%%%%%%%%%%%%%%%%%%%%%%%%%%%%%%%%%%%%%%%%%%%%%%%%%%%%%%%%%%%%%%%%%
Here we note that, as discussed in the previous section, the spin
state within each sublattice is ferromagnetic. This means that not
only must the total spin quantum numbers of $\S_A$ and $\S_B$ take their
maximum values, but the total spin quantum numbers of $\S_{A_1},\
\S_{A_2},\ \S_{B_1}$ and $\S_{B_2}$ must also take their maximum
values. More importantly, these values are thus fixed, which enables
us to treat the operators $\S_{A_1},\ \S_{A_2},\ \S_{B_1}$ and
$\S_{B_2}$ as four single spins, and in what follows we shall denote
these operators by their corresponding spin quantum numbers. The
problem is then that we are given a four spin state in which the spins
$S_{A_1}$ and $S_{A_2}$ are combined into a state with total spin
$S_A$ and the spins $S_{B_1}$ and $S_{B_2}$ are combined in a state
with total spin $S_B$, and then these two states are combined into a
total singlet (resulting in the ground state of our long-range AFM
model), and we must express this state in a basis in which the spins
$S_{A_1}$ and $S_{B_1}$ are combined into a state with total spin
$S_1$ and $S_{A_2}$ and $S_{B_2}$ are combined into a state with total
spin $S_2$.  This change of basis involves the familiar LS-jj coupling
scheme.

%Another consequence is that we
%no longer need to specify them when denoting a state formed by
%these four spins, and in order to keep things simple, we will use
%this convention for the rest of this paper.
%%%%%%%%%%%%%%%%%%%%%%%%%%%%%%%%%%%%%%%%%%%%%%%%%%%%%%%%%%%%%%%%%

We proceed by first rewriting the density
matrix describing the ground state in terms of the basis states $|S_1 m_1 S_2 m_2 \ket$ by appropriate insertions of
the unit operator $$\sum_{S_1 m_1 S_2 m_2}|S_1 m_1 S_2 m_2 \ket \bra
S_1 m_1 S_2 m_2|,$$
\beq\label{eqn:rdm}
\begin{split}
%%%%%%%%%%%%%%%%%%%%%%%%%%%%
%%% typo corrected below %%%
%%%%%%%%%%%%%%%%%%%%%%%%%%%%
  \rho &= |S_AS_B;Sm\ket \bra S_AS_B;Sm| = |\frac{N}{2} \frac{N}{2} ;
  00 \ket \bra \frac{N}{2}\frac{N}{2};00| \\
&= \sum_{\substack{S_1 m_1\\ S_2 m_2}}|S_1 m_1 S_2 m_2 \ket \bra S_1 m_1 S_2
  m_2|S_AS_B;Sm\ket  \\ & \times \bra S_AS_B;Sm| \sum_{\substack{S_1' m_1'\\ S_2' m_2'}}|S_1'
  m_1' S_2' m_2' \ket \bra S_1' m_1' S_2' m_2'|\\
  &= \sum_{\substack{S_1 m_1 S_2 m_2 \\ S_1' m_1' S_2' m_2'}} \lambda_{S_1 m_1 S_2
  m_2} \lambda_{S_1' m_1' S_2' m_2'}^*|S_1 m_1 S_2 m_2 \ket \bra S_1' m_1' S_2' m_2'|,\\
  %&= \sum_{S_1 m_1} \lambda_{S_1 m_1}|S_1 m_1 S_1 -m_1 \ket \bra S_1 m_1 S_1 -m_1|\\
\end{split}
\eeq
where $\lambda_{S_1 m_1 S_2 m_2} = \bra S_1 m_1 S_2
m_2|S_AS_B;Sm\ket$. We then insert another unit operator
$$\sum_{S_0 M}|S_1 S_2;S_0 M\ket\bra S_1 S_2;S_0 M|$$
to express this matrix element as the product of a Clebsch-Gordan
coefficient and an LS-jj coupling coefficient\cite{Dev},
\begin{equation}
\begin{split}
& \bra S_1 m_1 S_2 m_2|S_A S_B; S\ m\ket \\
& = \sum_{S_0, M} \bra S_1 m_1 S_2 m_2|S_1 S_2 S_0 M\ket\bra S_1 S_2 S_0 M|S_A S_B  S m\ket \\
& = \delta_{S_0 S}\delta_{m_1+m_2,m} \delta_{M, m} \bra S_1 m_1 S_2 m_2|S_1
  S_2 Sm \ket \begin{bmatrix} S_{A_1} & S_{B_1} & S_1\\ S_{A_2} & S_{B_2} & S_2\\ S_A & S_B & S\\\end{bmatrix}.
\end{split}
\end{equation}
%%%%%%%%%%%%%%%%%%%%%%%%%%%%%%
Here $\begin{bmatrix} S_{A_1} & S_{B_1} & S_1\\ S_{A_2} & S_{B_2} &
  S_2\\ S_A & S_B & S\\\end{bmatrix}$ is the LS-jj coupling
coefficient or X-coefficient defined as,
\beq
\begin{bmatrix} l_1 & s_1 & j_1\\ l_2 & s_2 & j_2\\
  L & S & J\\\end{bmatrix} = \bra l_1 l_2 s_1 s_2 L S J M | l_1 s_1
  l_2 s_2 j_1 j_2 J M\ket.
\eeq

% and is related to Wigner 9-j symbol
%  \cite{messiah} by the following equation:
%\begin{equation}
%\begin{split}
%& \begin{bmatrix} S_{A_1} & S_{B_1} & S_1\\ S_{A_2} & S_{B_2} & S_{2}\\
%  S_A & S_B & S\\\end{bmatrix} = \sqrt{(2S_1 + 1)(2S_2 + 1)}\\ &\times
%  \sqrt{(2 S_A + 1)(2 S_B + 1)}\begin{pmatrix} S_{A_1} & S_{B_1} & S_1\\ S_{A_2} & S_{B_2} & S_{2}\\ S_A & S_B & S\\\end{pmatrix}_{9j}
%\end{split}
%\end{equation}
Since $S = m = 0$, we have $S_1 = S_2,\ m_1 = -m_2$. So $\bra
S_1 m_1;S_{2} m_{2}|S_1 S_{2};S m\ket = \frac{(-1)^{S_1 - m_1}}{\sqrt{2 S_2 +
1}}$, and for simplicity, we can now suppress the $S_2$ and $m_2$ indices and
represent $\lambda_{S_1m_1S_2m_2}$ by $\lambda_{S_1m_1}$ without
causing any ambiguity. Therefore,
\begin{equation}\label{eqn:lsjj}
\begin{split}
&\lambda_{S_1m_1} = \frac{(-1)^{S_1 -
      m_1}}{\sqrt{2S_{1}+1}}\begin{bmatrix} S_{A_1} & S_{B_1} &
    S_{1}\\ S_{A_2} & S_{B_2} & S_{1}\\ S_A & S_B & S\\\end{bmatrix}\\ 
%&= \sqrt{(2S_{2}+1)(2S_A+1)(2S_B+1)}\begin{pmatrix} S_{A_1} & S_{B_1} & S_{2}\\ S_{A_2} & S_{B_2} & S_{_2}\\ S_A & S_B & S\\\end{pmatrix}_{9j}.
\end{split}.
\end{equation}
In the following calculation, for consistency, we adopt the
convention of Wigner 6-j and 9-j symbols, the definitions of
which, and the relation to the Racah coefficients and LS-jj
coupling coefficients (or X-coefficients), are given in Appendix
A.

Following Wigner's convention, Eq.~(\ref{eqn:lsjj}) can be written as,
\begin{equation}
\begin{split}
 \lambda_{S_1m_1} =& (-1)^{S_1 -
 m_1}\sqrt{(2S_{1}+1)(2S_A+1)(2S_B+1)} \\
 &\times \begin{pmatrix} S_{A_1} &
 S_{B_1} & S_{1}\\ S_{A_2} & S_{B_2} & S_{1}\\ S_A & S_B &
 S\\\end{pmatrix}.
\end{split}
\end{equation}

In our case $S=0$, so the 9-j symbol can be expressed more simply
in terms of a 6-j symbol\cite{BrinkA3} (see also Appendix B),
\begin{equation}\label{eqn:SZero}
\begin{split}
\begin{pmatrix} a & b & c\\ d & e & f \\ g & h & 0\\\end{pmatrix} & =
  \delta_{cf}\delta_{gh}(-1)^{a+d+c+g}((2c+1))^{-\frac{1}{2}}\\
& \times (2g+1)^{-\frac{1}{2}}\begin{pmatrix} a & b & c \\ e & d & g \end{pmatrix}.
\end{split}
\end{equation}

In our particular case, within each sublattice, the state is
ferromagnetic, $S_A=\frac{N}{2}, S_B=\frac{N}{2}$. Let $2N_1$ be
the size (number of lattice sites) of subsystem 1. Without losing
any generality, we can let $N_1 \leqslant N$. Applying Eq.
(\ref{eqn:SZero}) we then have,
\begin{equation}
  \lambda_{S_1,m_1} = (-1)^{S_1 - m_1}\sqrt{(N+1)}\begin{pmatrix} \frac{N-N_1}{2} &
  \frac{N-N_1}{2} & S_1 \\ \frac{N_1}{2} & \frac{N_1}{2} & \frac{N}{2} \end{pmatrix}.
  \label{schmidt}\\
\end{equation}
Using a symmetry property of the Racah coefficients\cite{BrinkA2}
$W(abcd;ef) = W(acbd;fe)$, we have,
\begin{equation}
 \lambda_{S_1m_1} = (-1)^{S_1 - m_1}\sqrt{(N+1)}\begin{pmatrix}
 \frac{N-N_1}{2} & \frac{N_1}{2} & \frac{N}{2} \\ \frac{N_1}{2} &
 \frac{N-N_1}{2} & S_{1} \end{pmatrix}.
\end{equation}
Then we employ the following relation\cite{BrinkA2}:
\begin{equation}
\begin{split}
   &W(abcd,a+b,f) =
  \bigg[\frac{(2a)!(2b)!}{(2a+2b+1)!(c+d-a-b)!} \\
    & \times
  \frac{(a+b+c-d)!(a+b+d-c)!!}{(a+f-c)!(a+c+f+1)!(b+d-f)!}\\
  & \times
  \frac{(c+f-a)!(d+f-b)(a+b+c+d+1)!}{(b+f-d)!(b+f+d+1)!(a+c-f)!}\bigg] ^{\frac{1}{2}}.\\
\end{split}
\end{equation}
This gives
\begin{equation}\label{eqn:lambda}
\begin{split}
 \lambda_{S_{1},m_{1}} &=
 (-1)^{S_1 - m_1}\bigg[\frac{(N+1)(N-N_1)!N_1!}{(N-N_1-S_{1})!(N-N_1+S_{1}+1)!}\\
& \times \frac{(N-N_1)!N_1!}{(N_1-S_{1})!(N_1+S_{1}+1)!}\bigg]^\frac{1}{2}.
\end{split}
\end{equation}

Now with the expression above, we can directly trace out the
degrees of freedom in subsystem 2 from Eq. (\ref{eqn:rdm}), with
the result
\beq \rho_1 = tr_{(2)}(\rho) = \sum_{S_1 m_1} |\lambda_{S_1
m_1}|^2|S_1 m_1 \ket \bra S_1 m_1|. \eeq
The bipartite entanglement entropy between subsystems 1 and 2 is
then given by
\begin{equation}\label{eqn:entropy1}
E= E_1 = -\sum_{S_1,m_1} \abs{\lambda_{S_1 m_1}}^2\ln(\abs{\lambda_{S_1 m_1}}^2).
\end{equation}

Here we note that, although $\lambda_{S_1m_1}$ is written with an
explicit $m_1$ dependence, the actual expression is independent of
$m_1$. As a result, we can eliminate the summation over $m_1$ from
(\ref{eqn:entropy1}) by multiplying by a factor of $2S_1+1$.  The
final expression for the entanglement entropy is then
\begin{equation}\label{eqn:entropy}
E= E_1 = -\sum_{S_1}(2S_1 + 1) \abs{\lambda_{S_1 m_1}}^2\ln(\abs{\lambda_{S_1 m_1}}^2).
\end{equation}

In the following section we will first derive the asymptotic behavior of
the entanglement entropy $E$ in certain limits using the
exact results above, and then present the results of our
numerical calculation of $E$ with finite system sizes.

%%%%%%%%%%%%%%%%%%%%%%%%%%%%%%%%%%%%

%%%%%%%%%%%%%%%%%%%%%%%%%%%%%%%%%%%%
%                              %%%%%
%  Section IV Numerical Result %%%%%
%                              %%%%%
%%%%%%%%%%%%%%%%%%%%%%%%%%%%%%%%%%%%
%\section{NUMERICAL CALCULATION}

%%%%%%%%%%%%%%%%%%%%%%%%%%%%%%%%%%%%%%%
%                                 %%%%%
% SectionIII: Asymptotic Behavior %%%%%
%                                 %%%%%
%%%%%%%%%%%%%%%%%%%%%%%%%%%%%%%%%%%%%%%
\section{ASYMPTOTIC BEHAVIOR and entanglement entropy}
In this section we consider two limiting cases:

(i)$N_1 = N_2 = N$, this case gives the saturated entropy at fixed $N$ since
  intuitively $E$ should increase with the subsystem size.

(ii) $1 \ll N_1 \ll N$, in this limit we are considering
  system's entanglement with its (much larger) environment, and generically
we should be able to find that the entropy should be independent of
  the total system size as $N \rightarrow \infty$, which is indeed what we find.

To obtain the asymptotic behavior of the entanglement entropy, we
first consider the asymptotic behavior of $\lambda_{S_1m_1}$, then
in the large $N$ limit. Then the summation over $S_1$ which runs from 0
to $N/2$ can be replaced by an integral. As we will see, the
distribution of $\lambda_{S_1m_1}$ is proportional to a Gaussian
function with respect to $S_1$, thus the bounds of integration can
be extended to from 0 to $+\infty$.

%\subsection{Asymptotic Behavior of $\lambda_{S_1m_1}$}
Our concern is the distribution of $\lambda_{S_1m_1}$ with respect
to $S_1$, therefore, we extract the dependence on $S_1$ and then
use the normalization condition $\sum_{S_1m_1} \lambda_{S_1 m_1}^2
= \sum_{S_1} (2S_1 + 1) \lambda_{S_1m_1}^2 = 1$ to obtain the
normalization constant. From Eq. (\ref{eqn:lambda}) we can write,
\beq
\begin{split}
\lambda_{S_1m_1}^2 &\sim
\frac{(2N_1)!}{(N_1-S_1)!(N+S_1+1)!}\\
& \times \frac{(2N-2N_1)!}{(N - N_1 -  S_1)! (N - N_1 + S_1 +1)!}\\
&\sim e^{-\frac{S_1^2}{N_1} - \frac{S_1^2}{N - N_1}}.\\
\end{split}
\eeq
This approximation is generally good as long as both $N_1$ and $N
- N_1$ are large. In our work, $N \gg 1$, and $N_1$ is assumed to
be $\frac{N}{2}$ (equal partition) or $N \gg N_1 \gg 1$, so this
criteria is satisfied quite well in both cases. Then by the
normalization condition \beq
\begin{split}
  &\sum_{S_1}A(2S_1+1)e^{-S^2_{1}(\frac{1}{N-N_1}+\frac{1}{N_1})} \\
  &\simeq \int_0^\infty A(2S_1+1)e^{-S_1^2(\frac{1}{N - N_1}+\frac{1}{N_1})}dS_1 = 1,\\
\end{split}
\eeq
we obtain the (inverse) precoefficient  $\frac{1}{A} = (N_1 - \frac{N_1^2}{N} +
\frac{1}{2}\sqrt{\frac{\pi (N-N_1)N_1}{N}})$. Therefore,
\beq \lambda_{S_1m_1}^2 \simeq \frac{1}{N_1 - \frac{N_1^2}{N} +
\frac{1}{2}\sqrt{\frac{\pi (N-N_1)N_1}{N}}} e^{-\frac{S_1^2}{N_1}
-
  \frac{S_1^2}{N - N_1}}.
\eeq
The entanglement entropy is therefore given by
\begin{equation}
  \begin{split}
    E_1 \simeq & - \int_0^\infty
    A(2S_1+1)e^{-S^2_{1}(\frac{1}{N-N_1}+\frac{1}{N_1})} \\ & \times \ln \left(Ae^{-S^2_{1}(\frac{1}{N-N_1}+\frac{1}{N_1})}\right) dS_1 \\
    \simeq & \ln\left(N_1 - \frac{N_1^2}{N} + \frac{1}{2}\sqrt{\frac{\pi (N-N_1)N_1}{N}}\right).
  \end{split}
\end{equation}

%%%%%%%%%%%%%%%%%%%
%equal partition
%%%%%%%%%%%%%%%%%%%

\subsection{Equal Partition}
%%%%%%%% revision here 4.07%%%%%%%%%%
Here for simplicity we set $N$ even (note that the total system size is
$2N$), thus  $N_1 = N_2 = N/2$,
and $\frac{1}{A}=\frac{1}{4}(N+\sqrt{N\pi})$, then
\begin{equation}
\lambda_{S_1m_1}^2 = \frac{4}{N+\sqrt{N\pi}}
e^{-\frac{4S_1^2}{N}},
\end{equation}
and the entanglement entropy becomes,
\begin{equation}
\begin{split}
  E &= \ln \left(\frac{N}{4} + \frac{1}{4}\sqrt{\pi N} \right)\\
& \simeq \ln N -\ln 4 \simeq \ln N - 1.38629.\\
\end{split}
\end{equation}
Compared with our numerical result, as shown in Fig.~\ref{Fig.ep},
we see that they agree very well, not only for the prefactor of
the $\ln N$ dependence, but the intersection coefficient as well.

\bef
\includegraphics[width = 8cm]{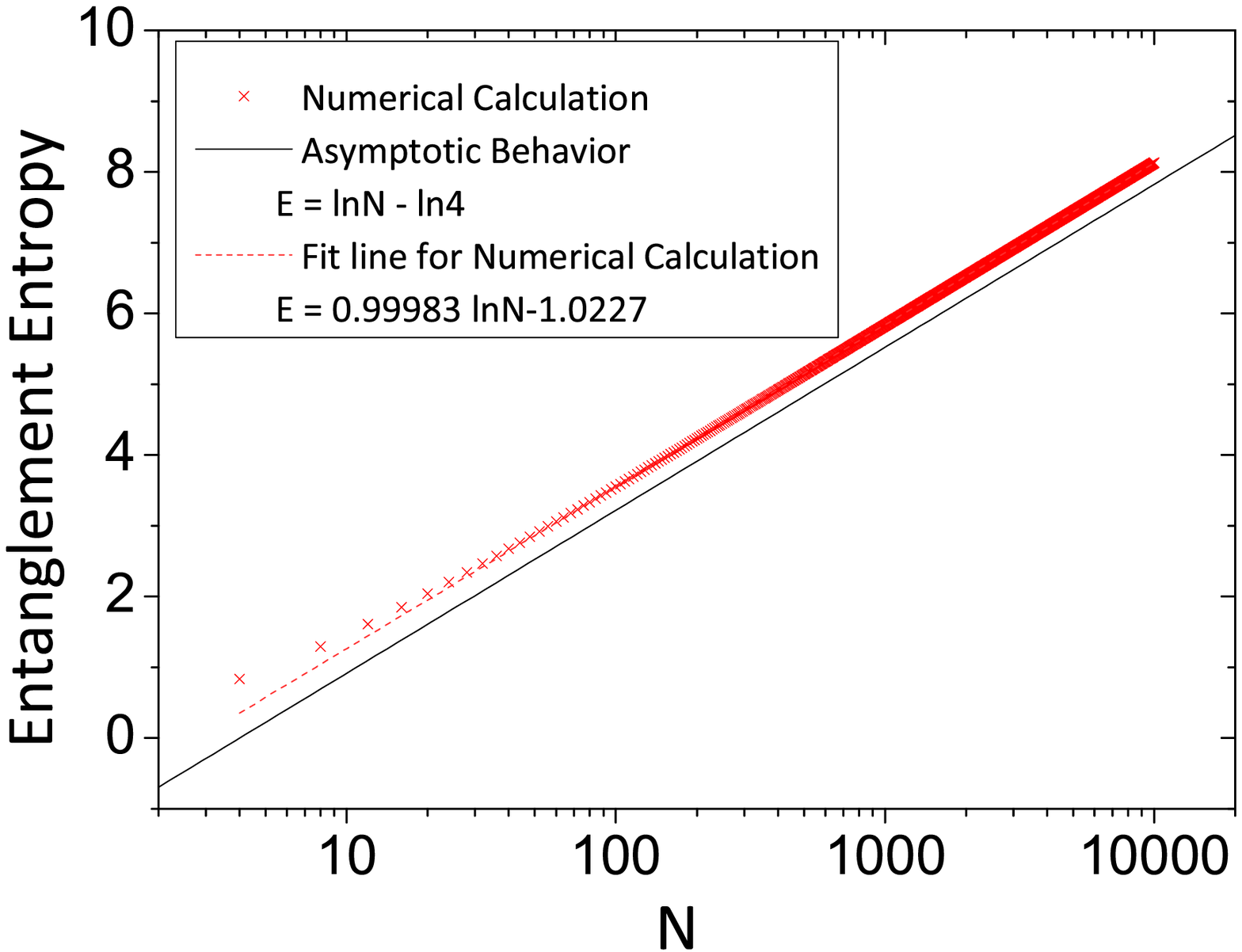}
\caption{(Color online) Comparison of asymptotic behavior (black solid line) and exact numerical
  calculation (scattered points) of entanglement entropy $E$. Exact
  calculation of $E$ for the equal partition case is compared with
  asymptotic behavior we derived in the text up to $N=10000$. Results
  obtained from these two methods agree well. Our approximation of the
  asymptotic behavior gives $E \simeq \ln(N) - 1.38629$, compared with
  the best fit line (red dashed line) $E = 0.99983 \ln N -
  1.0227$.}\label{Fig.ep} 
\enf

% subsection unequal partition %
\subsection{Unequal Partition}
In this case, $1 \ll N_1 \ll N$, we can expand the entropy as
follows,
\beq
\begin{split}
E &\simeq \ln \left[ N_1 \left(1 - \frac{N_1}{N} + \frac{1}{2}
\sqrt{\frac{\pi (1 -
    \frac{N_1}{N})}{N_1}}\right)\right]\\
& = \ln N_1 + \ln \left(1 - \frac{N_1}{N} + \frac{1}{2}
\sqrt{\frac{\pi (1 -
    \frac{N_1}{N})}{N_1}}\right)\\
& \simeq \ln N_1 - \frac{N_1}{N} +
O\left(\left(\frac{N_1}{N}\right)^2\right).
\end{split}
\eeq
From this expression we see that when the assumed condition is
satisfied the entropy indeed depends only on the subsystem size to
leading order. To check to what extent our approximation is still
valid, we plot the expression as a function of total system size
$N$ with different subsystem size $N_1$, as shown in Fig.
\ref{Fig.uep.N}. This figure shows that when $N$ is large enough
the entropy becomes independent of $N$, and the first order term,
$\ln N_1$ dominates except for an apparently constant shift from
the exact curve. We have also plotted the asymptotic behavior as a
function of $N_1$ while keeping $N$ fixed, shown in
Fig.\ref{Fig.uep.l}.

\bef
\includegraphics[width=8cm]{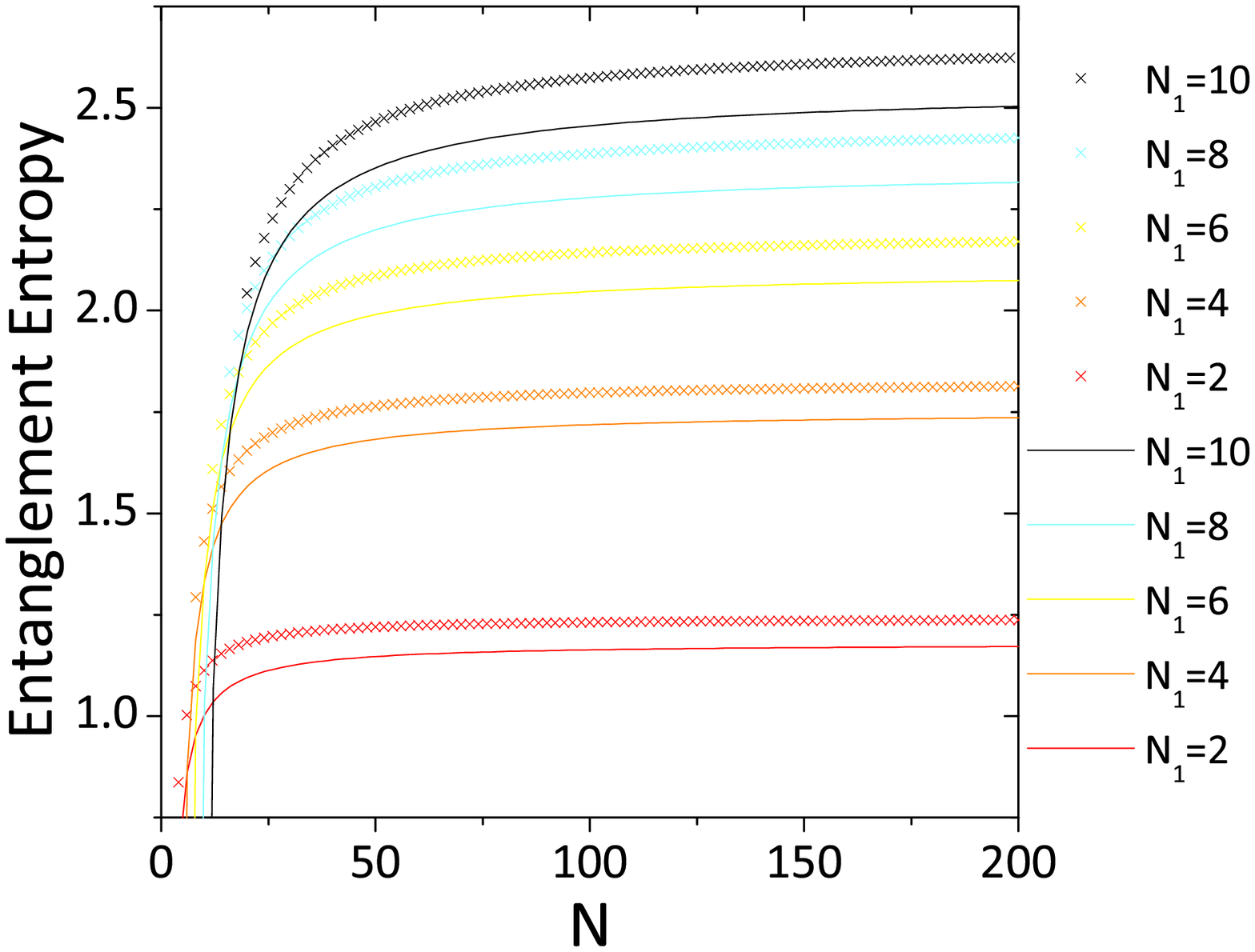}
\caption{(Color online) Comparison of asymptotic behavior (solid line) and
  exact numerical calculation (scattered points) of entanglement
  entropy $E$. As total system size $N$ increases, $E$ tends to a
  constant for a fixed subsystem size $N_1$, as expected. Difference
  between them increases as the subsystem size increases, as the
  condition $1 \ll N_1 \ll N/2$ gets less satisfied.}\label{Fig.uep.N}
\enf

\bef
\includegraphics[width=8cm]{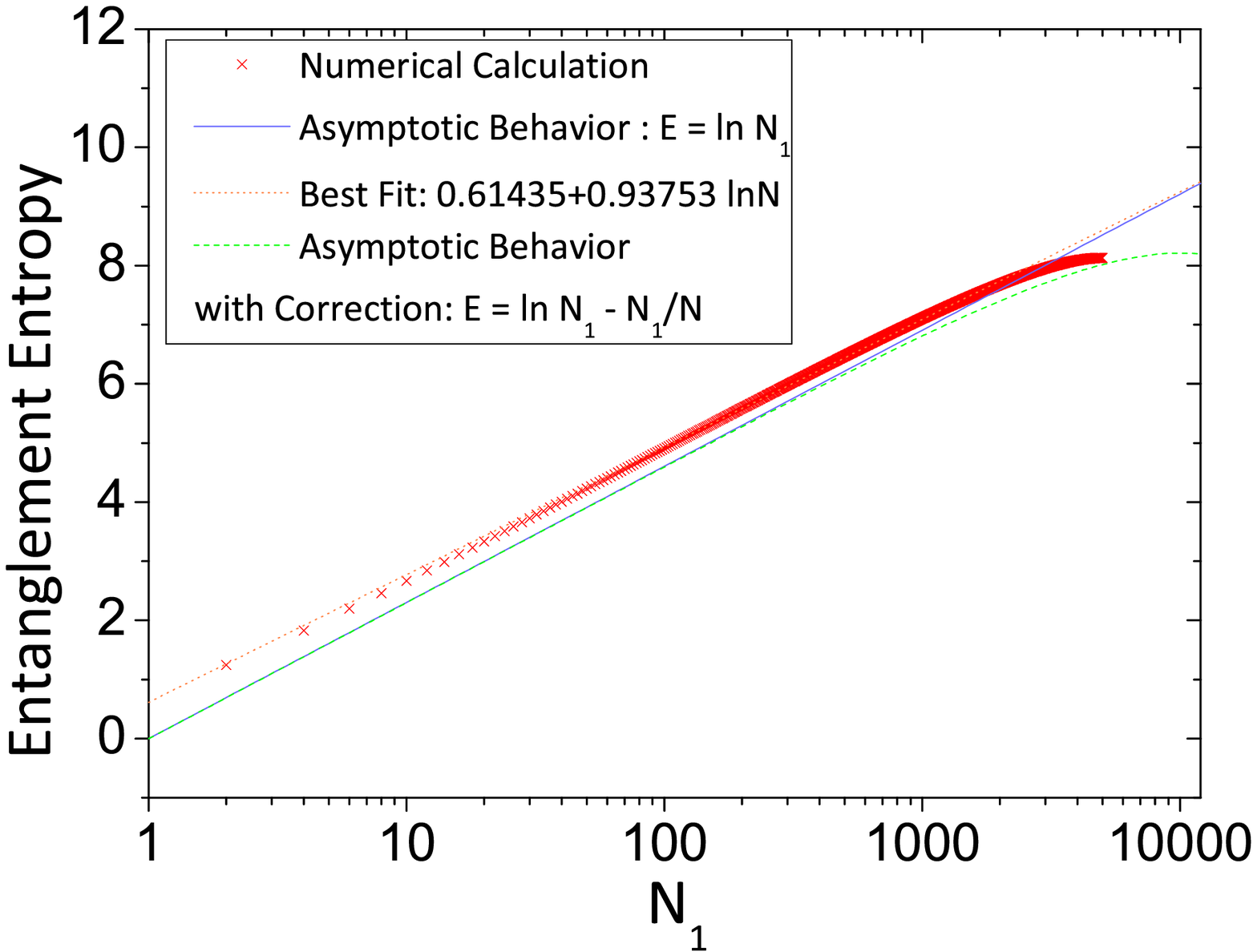}
\caption{(Color online) Asymptotic behavior of entanglement entropy $E$ as function of $N_1$ with
  $N$ fixed (solid blue line). Here $N = 10000$. Compare it with the
  scattered points, which are an exact numerical calculation of $E$. We
  find the asymptotic behavior is accurate until $N_1$ becomes
  comparable to $N$. We also plot the asymptotic behavior with first
  order correction(dashed light green line) which exhibits the correct
  tail effect when $N_1$ becomes comparable with $N$.}\label{Fig.uep.l}
\enf

\section{Ferromagnetic model and its entanglement entropy}

In this section we consider a ferromagnetic (FM) spin-1/2 model on
%%%%%%%%%%%%%%%%%%%%%%%%%%%%%%%%
% newly added, specify N even %%
%%%%%%%%%%%%%%%%%%%%%%%%%%%%%%%%
an arbitrary lattice with $N$ sites,
\beq H=-\sum_{i\ne
j}J_{ij}\S_i\cdot\S_j, \eeq
with $J_{ij} > 0$. The ground state is the fully magnetized state
$|SM\rangle$ with $S=N/2$ and $M=-S, -S+1, \cdots, S$, and is
clearly long-range ordered: $\langle\S_i\cdot\S_j\rangle=1/4$.
However there is a crucial difference between the FM ground state
and the AFM ground state studied earlier: the FM ground state has
a finite degeneracy, and thus the system exhibits a non-zero
entropy even at zero temperature, $E_{0} =\log(2S+1)=\log(N+1)$,
resulting from the density matrix of the entire system,
\beq
\begin{split}
\rho &=\frac{1}{2S+1}\sum_{M=-S}^S | SM \ket \bra SM | \\
& = \frac{1}{N + 1} \sum_{M=-S}^S | SM \ket \bra SM | \\
& = \frac{1}{N + 1}     \left(  \begin{array} {ccccc}
      1 & 0 &  & \dots & \\
      0 & 1 & 0 &  & \\
      \vdots  & 0 & 1 & 0 & \\
        &  &  & \ddots & \\
        &  &\dots & 0 & 1
  \end{array}  \right)\\
\end{split}
\eeq

In this case the entanglement entropy between two subsystems (1 and 2)
is defined in the following manner. We first obtain reduced density
matrices for subsystems 1 and 2 by tracing out degrees of freedom in
2 and 1 from $\rho$:
\beq
\begin{split}
&\rho_1=tr_{(2)}\rho = \frac{1}{N+1} \sum_{M = -S}^{S} tr_{(2)} (| SM \ket
\bra SM |) \\
&= \frac{1}{N+1} \sum_{M = -S}^{S} \rho_{1M},\\
&\rho_2= tr_{(1)}\rho = \frac{1}{N+1} \sum_{M = -S}^{S} tr_{(1)} (| SM
\ket \bra SM |)\\
&= \frac{1}{N+1} \sum_{M = -S}^{S} \rho_{2M},
\end{split}
\eeq and calculate from them the entropy of the subsystems, $E_1$ and
$E_{2}$. The entanglement entropy is defined
as\cite{entanglenote} 
\beq E
=(E_{1} + E_{2} - E_{0}) / 2. 
\eeq

For the present case $E_{1}$ and $E_{2}$ can be easily obtained
from the following observations. (i) Because the total spin is
fully magnetized, so are those in the subsystems: $S_1=N_1/2$ and
$S_2=N_2/2$. Thus this is a two-spin entanglement problem. (ii)
Because the total density matrix $\rho$ is proportional to the
identity matrix in the ground state subspace, it is invariant
under an arbitrary rotation in this subspace. (iii) As a result
the reduced density matrix $\rho_1$ is also invariant under
rotation in the subspace of subsystem 1 with $S_1=N_1/2$, and is
proportional to the identity matrix in this subspace. Thus \beq
\begin{split}
\rho_1 &= \frac{1}{N_1 + 1} \sum_{M_1=-\frac{N_1}{2}}^\frac{N_1}{2} |
\frac{N_1}{2}M_1 \ket \bra \frac{N_1}{2}M_1 | \\
& =\frac{1}{N_1 + 1}     \left(  \begin{array} {ccccc}
      1 & 0 &  & \dots & \\
      0 & 1 & 0 &  & \\
       & 0 & 1 & 0 & \\
      \vdots  &  &  & \ddots & \\
        &  &\dots & 0 & 1
  \end{array}  \right),\\
\end{split}
\eeq
and $E_{1}=\ln(N_1+1)$ (in agreement with Ref.
\cite{popkev}). Similarly $E_{2}=\ln(N_2+1)$. Thus
\beq E =
[\ln(N_1+1)+\ln(N_2+1)-\ln(N+1)]/2.
\eeq
We find in both the equal partition ($N_1=N_2=N/2$, here again we set
$N$ even for simplicity) and unequal
partition ($N_1\ll N_2=N-N_1$) limits, the entropy grows
logarithmically with subsystem size $N_1$,
\beq \lim_{N_1\rightarrow\infty}E \rightarrow {1\over 2}\ln(N_1).
\eeq

%%%%%%%%%%%%%%%%%%%%%%%%%
%                   %%%%%
%    Conclusion     %%%%%
%                   %%%%%
%%%%%%%%%%%%%%%%%%%%%%%%%
\section{Concluding remarks}

In this paper we have studied antiferromagnetic (AFM) and
ferromagnetic (FM) spin models with perfect long-range magnetic
order in their ground states, and calculated the ground state block
entanglement entropy when the system is divided into two subsystems
(or blocks). In both cases we find the entropy grows logarithmically
with block size. In the following we discuss the significance of our
results.

First of all, there {\em is} entanglement, despite the perfect
long-range order in the ground state. This is somewhat surprising
as one might think that in such Hamiltonians one can obtain
certain properties of the system exactly using mean-field theory,
in which the ground state is approximated by a product state with
no entanglement. Our results indicate that mean-field
approximation is {\em not} appropriate for entanglement
calculation, even if it is ``exact" for other purposes. This point
is particularly striking for the AFM model, in which the ground
state is unique. The source of the discrepancy is there still is
quantum fluctuations even in such a model with super long-range
interaction, which render the ground state a singlet, even though
quantum fluctuation does not reduce the size of the order
parameter. The entanglement is due to the quantum fluctuation of
the {\em direction} of the order parameter, which is a collective
mode with zero wave-vector (or a zero mode); this is missed by any
mean-field approximation.

Secondly, the entropy does {\em not} obey the area law. The
reasons for that are different for the two cases we studied. In
the AFM model, the interaction does not depend on distance in the
Hamiltonian, thus there is no notion of distance or area in this
model. In the FM model, on the other hand, the ground state is
{\em independent} of the Hamiltonian, as long as all interactions
are FM. The fully magnetized ground states are invariant under
permutation of spins, thus there is no notion of distance or area
in the {\em ground states}, even though the terms in Hamiltonian
can have distance dependence.

Thirdly, the absence of an area law is special to the cases we
studied, again each in their own ways. For the FM model, it is
specific to zero temperature. As soon as a finite temperature is
turned on, one expects the entanglement entropy (or mutual
information) to grow with the area separating the two subsystems
or blocks, as long as the interaction is not
long-ranged\cite{wolf07}. For the AFM model at zero temperature,
we do expect an area-law contribution to the entropy for
short-range or even certain power-law long-ranged interaction due
to quantum fluctuations. This is most easily seen within spin-wave
approximation, which is a version of mean-field theory. Within the
spin wave approximation spins are mapped onto bosons, and the
Hamiltonian is mapped onto coupled harmonic oscillators. Detailed
recent studies have established the area law of entanglement for
such systems\cite{amico}. In this regard the super long-range AFM
model we study here is very special, in that all spin-wave degrees
of freedom at finite wave-vector disappear, and the only degrees
of freedom contributing to the ground state are zero wave-vector
modes represented by ${\bf S}_{A}$ and ${\bf S}_{B}$\cite{yusuf};
as a result there is no area-law contribution from quantum
fluctuations of spin-waves.

We conclude by speculating that in the more general cases that do
have an area-law contribution to the entanglement entropy, as long
as long-range spin order is present, the logarithmic contribution
due to fluctuations of the order parameter zero-modes we find here
will remain and show up as a sub-leading (yet singular) correction
to the area law. If that is the case, then conventional long-range
order contributes to the entanglement entropy in a way similar to
the much subtler topological order\cite{kitaevpreskill,levinwen}
or quantum criticality\cite{fradkinmoore}.

\acknowledgments

We acknowledge support from National Science Foundation grants
No. DMR-0225698 and No. DMR-0704133 (W.D. and K.Y.), State of Florida (W.D.),
 and US DOE Grant No. DE-FG02-97ER45639 (N.E.B.).
%%%%%%%%%%%%%%%%%%%%%%%%%
%                   %%%%%
%     Appendix      %%%%%
%                   %%%%%
%%%%%%%%%%%%%%%%%%%%%%%%%
\appendix

\section{}

6-j Symbol \& Racah coefficient:
\begin{equation}
\begin{split}
   &\begin{pmatrix} j_1 & j_2 & J_{12} \\ j_3 & J & J_{23}
   \end{pmatrix} = (-)^{j_1+j_2+j_3+J}W(j_1j_2Jj_3;J_{12}J_{23}) \\
&= (-1)^{j_1+j_2+j_3+J}[(2J_{12}+1)(2J_{23}+1)]^{-\frac{1}{2}}\\
& \times \bra j_1,J_{23};J|J_{12},j_3;J\ket.\\
\end{split}
\end{equation}
where $J_{12}$ and $J_{23}$ refer to the coupling of $j_1$ and $j_2$
or $j_2$ and $j_3$ respectively.

9-j Symbol \& LS-jj Coupling coefficient:
\begin{equation}
  \begin{split}
    & \begin{pmatrix} a & b & c\\ d & e & f \\ g & h & i\\\end{pmatrix} =
      ((2c + 1)(2f + 1)(2g + 1))^{-\frac{1}{2}}\\
      & \times (2h + 1)^{-\frac{1}{2}} \begin{bmatrix} a & b & c\\ d & e & f \\ g & h & i\\\end{bmatrix}\\
    & = ((2c + 1)(2f + 1)(2g + 1)(2h + 1))^{-\frac{1}{2}} \\
    &\ \ \  \times \bra (ab)c,(de)f;i|(ad)g,(bc)h;i\ket.
  \end{split}
\end{equation}

\section{}

In this Appendix we give a slightly simplified derivation of the
entanglement entropy for the singlet ground state of the
infinite-range AFM model defined in Sec.~II.  While this
derivation is not as general as that given in the main text (which
can, in principle, be applied to states with nonzero total spin),
it has the advantage of clarifying the reason for the appearance
of the Wigner 6j-symbol in the final result for the entanglement
entropy.

Following the notation of Sec. II and III, we consider the case of
$2N$ spin-1/2 particles divided equally into two sublattices $A$
and $B$. The spins on each sublattice are taken to be fully
polarized, so that $S_A = N/2$ and $S_B = N/2$.  In what follows
we will use a notation in which, for example, the state for which
the spins $S_A$ and $S_B$ form a singlet is represented as
$(S_A,S_B)_0$. In this notation, pairs of spins contained within a
set of parenthesis form a state whose total spin is equal to the
subscript labeling the parenthesis. Thus,
$\left(S_{A},S_{B}\right)_0$ is a singlet state equivalent to the
state defined in Eq.~(\ref{singletneel}) in Sec.~II.  (Needless to
say, for this state to exist it is necessary to have $S_A = S_B$.)

Now we consider what happens if these spins are divided into two
different subsystems labeled $1$ and $2$.  Following Sec.~II, if
$S_{A_i}$ is the total spin quantum number of the $A$ sublattice
spins in subsystem $i=1,2$, and $S_{B_i}$ is the total spin
quantum number of the $B$ sublattice spins in subsystem $i=1,2$,
then the state of the spins on the $A$ sublattice can be written
$(S_{A_1},S_{A_2})_{S_A}$ and the state of the spins on the $B$
sublattice can be written $(S_{B_1},S_{B_2})_{S_B}$.  The total
singlet state for the entire system, $|\psi\rangle$, can then be
expressed as
\begin{eqnarray}
|\psi\rangle =
\left(\left(S_{A_1},S_{A_2}\right)_{S_A},
\left(S_{B_1},S_{B_2}\right)_{S_B}\right)_0.
\label{singlet}
\end{eqnarray}

Note that we have not included the $m$ quantum numbers associated
with the $z$-components of $S_A$ and $S_B$  in writing the above
expression for $|\psi\rangle$.  This is not necessary because the
requirement that the total spin of the two sublattices combine to
form a singlet uniquely determines the state. It should, of
course, always be understood that when we write
Eq.~(\ref{singlet}) what is really meant (in obvious notation) is,
\begin{eqnarray}
|\psi\rangle &=& \frac{1}{\sqrt{2S_A+1}} \sum_{m=-S_A}^{S_A}
(-1)^m|\left(S_{A_1},S_{A_2}\right)_{S_A};
m\rangle\nonumber\\
&&~~~\otimes |\left(S_{B_1},S_{B_2}\right)_{S_B};- m \rangle.
\label{trivial}
\end{eqnarray}

Equation (\ref{trivial}) effectively (up to irrelevant phase
factors) gives the Schmidt decomposition (see, e.g.,
\cite{nielsenchuang}) of the state $|\psi\rangle$ into the two
subsystems consisting of the $A$ and $B$ sublattices.  Given such
a Schmidt decomposition it is straightforward to determine the
entanglement between these two subsystems. However, here we are
interested not in the trivial entanglement between subsystems $A$
and $B$, but the entanglement between subsystems 1 and 2.  To find
this we need the Schmidt decomposition of $|\psi\rangle$ into
subsystems 1 and 2.

Note that because $|\psi\rangle$ is a total spin singlet it
follows that
\begin{eqnarray}
|\psi\rangle &=&
\left(\left(S_{A_1},S_{A_2}\right)_{S_A},\left(S_{B_1},S_{B_2}
\right)_{S_B}\right)_0\nonumber\\
&=& \left(\left((S_{A_1},S_{A_2})_{S_A},S_{B_2}\right)_{S_{B_1}},
S_{B_1}\right)_0,
\end{eqnarray}
where, for convenience, we have rearranged the order of the spins
in the second equality.  The key point here is that the three
spins $S_{A_1}$, $S_{A_2}$ and $S_{B_2}$ must be in a state which
is a total spin eigenstate with total spin $S_{B_1}$.  If this
were not the case it would be impossible to form a total spin
singlet with the remaining spin $S_{B_1}$.

Next we use the Wigner $6j$-symbol to express the three spin state
$\left((S_{A_1},S_{A_2})_{S_A},S_{B_2}\right)_{S_{B_1}}$ as a
superposition of states of the form
$\left(S_{A_1}\left(S_{A_2},S_{B_2}\right)_{S_{2}}\right)_{S_{B_1}}$.
The result is
\begin{eqnarray}
&&\left(\left(S_{A_1},S_{A_2}\right)_{S_A},{S_{B_2}}\right)_{S_{B_1}}\nonumber
\\
&&~~~~ = (-1)^N \sum_{S_2}  \gamma_{S_2}
 \left(S_{A_1},\left(S_{A_2}
 {S_{B_2}}\right)_{S_{2}}\right)_{S_{B_1}},
\end{eqnarray}
where the coefficients are given by
\begin{eqnarray}
\gamma_{S_2} = {\sqrt{(2S_A + 1)(2S_2 +1)}}
\left(\begin{array}{ccc} S_{A_1} & S_{A_2} & S_A \\ S_{B_2} &
S_{B_1} & S_2 \end{array}\right), \label{gamma}
\end{eqnarray}
(see Appendix A).

Rearranging the spins again, and using the fact that the total
spin of all four spins is 0, we can express the resulting four
spin basis states as follows,
\begin{eqnarray}
&&\left(\left(S_{A_1} \left(S_{A_2}, S_{B_2}\right)_{S_2}\right)_{S_{B_1}}, S_{B_1}\right)_0\nonumber\\
&&~~~~~~~~~~~~~ =
\left(\left(S_{A_1},S_{B_1}\right)_{S_1}, \left(S_{A_2},S_{B_2}\right)_{S_2}\right)_0
\end{eqnarray}
where, obviously, $S_1 = S_2$. We therefore conclude that
\begin{eqnarray}
|\psi \rangle = (-1)^N \sum_{S_2} \gamma_{S_2}
\left(\left(S_{A_1},S_{B_1}\right)_{S_2},\left(S_{A_2},S_{B_2}\right)_{S_2}\right)_0.
\end{eqnarray}
Finally, after writing the $m$ sum explicitly (as in
Eq.~(\ref{trivial})), we have the desired Schmidt decomposition
(again, up to irrelevant phases) of the state $|\psi\rangle$ into
subsystems 1 and 2,
\begin{eqnarray}
|\psi\rangle &=& (-1)^N \sum_{S} \sum_{m = -S}^S (-1)^m
\lambda_{Sm} |\left(S_{A_1},S_{B_1}\right)_{S},
m\rangle\nonumber\\
&&~~~\otimes |\left(S_{B_2},S_{A_2}\right)_{S}, - m \rangle
\end{eqnarray}
where
\begin{eqnarray}
\lambda_{Sm} = \sqrt{2 S_A + 1} \left(\begin{array}{ccc} S_{A_1} &
S_{A_2} & S_A \\ S_{B_2} & S_{B_1} & S_2
\end{array}\right).
\end{eqnarray}
When the values of $S_A = N/2$, $S_{A_1} = S_{B_1} = N_1/2$,
$S_{A_2} = S_{B_2} = (N-N_1)/2$ are substituted into this
expression it can be seen to be equivalent to that given in
Eq.~(\ref{schmidt}) in Sec.~III.  The derivation of the
entanglement entropy given in the main text follows.

%%%%%%%%%%%%%%%%%%%%%%%%%
%                   %%%%%
%     Reference     %%%%%
%                   %%%%%
%%%%%%%%%%%%%%%%%%%%%%%%%

%\newpage


\begin{thebibliography}{99}
\bibitem{note} The entanglement property of this model has bee previosuly studied in Ref.\onlinecite{popkev}. Here we introduce a different definition of the entanglement entropy that properly takes into account the ground state degeneracy, and can be calcualted much more straigtforwardly using symmetry consideration. See Sec. V.
\end{thebibliography}
\end{document}